\documentclass[prb,twocolumn,showpacs,floatfix,amsmath,amssymb,superscriptaddress]{revtex4}
\usepackage[dvips]{graphicx}
\usepackage{latexsym}
\usepackage{graphicx}
\usepackage{subfigure}
\usepackage{times}
\usepackage{amsmath}
\usepackage{dcolumn}
\usepackage{latexsym,amsmath,amssymb,bm,euscript}
\begin{document}

\title{Majorana fermions in a tunable semiconductor device}

\author{Jason Alicea}
\affiliation{Department of Physics, California Institute of Technology,
Pasadena, California 91125}

\date{\today}

\begin{abstract}
The experimental realization of Majorana fermions presents an important problem due to their non-Abelian nature and potential exploitation for topological quantum computation.  Very recently Sau \emph{et al}.\ [arXiv:0907.2239] demonstrated that a topological superconducting phase supporting Majorana fermions can be realized using surprisingly conventional building blocks: a semiconductor quantum well coupled to an $s$-wave superconductor and a ferromagnetic insulator.  Here we propose an alternative setup, wherein a topological superconducting phase is driven by applying an in-plane magnetic field to a (110)-grown semiconductor coupled \emph{only} to an $s$-wave superconductor.  This device offers a number of advantages, notably a simpler architecture and the ability to tune across a quantum phase transition into the topological superconducting state, while still largely avoiding unwanted orbital effects.  Experimental feasibility of both setups is discussed in some detail.

\end{abstract}

\maketitle

%%%%%%%%%%%%%%%%%%%%%%%%%%%%%%%%%%%%%%%%%%%%%%%%%%%%%%%%%%%%%%%%%%%%%%

\section{Introduction}

The problem of realizing and manipulating Majorana fermions in condensed matter systems is currently a topic of great theoretical and experimental interest.  Roughly, Majorana fermions constitute `half' of a usual fermion.  That is, creating an ordinary fermion $f$ requires superposing two Majorana modes $\gamma_{1,2}$---which can be separated by arbitrary distances---via $f = \gamma_1 + i \gamma_2$.  The presence of $2n$ well-separated Majorana bound states thus allows for the construction of $n$ ordinary fermions, producing (ideally) a manifold of $2^n$ degenerate states.  Braiding Majorana fermions around one another produces not just a phase factor, as in the case of conventional bosons or fermions, but rather transforms the state nontrivially inside of this degenerate manifold: their exchange statistics is non-Abelian\cite{ReadGreen,Ivanov}.  Quantum information encoded in this subspace can thus be manipulated by such braiding operations, providing a method for decoherence-free topological quantum computation\cite{Kitaev,TQCreview}.  Majorana fermions are therefore clearly of great fundamental as well as practical interest.

At present, there is certainly no dearth of proposals for realizing Majorana fermions.  Settings as diverse as fractional quantum Hall systems\cite{ReadGreen} at filling $\nu = 5/2$, strontium ruthenate thin films\cite{SrRu}, cold atomic gases\cite{pwaveColdAtoms,TQCcoldatoms,TopologicalSF,FujimotoColdAtoms}, superfluid He-3\cite{MajoranaHe3}, the surface of a topological insulator\cite{FuKane}, semiconductor heterostructures\cite{Sau}, and non-centrosymmetric superconductors\cite{SatoFujimoto,PatrickProposal} have all been theoretically predicted to host Majorana bound states under suitable conditions.  Nevertheless, their unambiguous detection remains an outstanding problem, although there has been recent progress in this direction in quantum Hall systems\cite{Willett1,Willett2}.  

Part of the experimental challenge stems from the fact that stabilizing topological phases supporting Majorana fermions can involve significant engineering obstacles and/or extreme conditions such as ultra-low temperatures, ultra-clean samples, and high magnetic fields in the case of the $\nu = 5/2$ fractional quantum Hall effect.  The proposal by Fu and Kane\cite{FuKane} noted above for realizing a topological superconducting state by depositing a conventional $s$-wave superconductor on a three-dimensional topological insulator surface appears quite promising in this regard.  This setting should in principle allow for a rather robust topological superconducting phase to be created without such extreme conditions, although experiments demonstrating this await development.  Moreover, Fu and Kane proposed methods in such a setup for creating and manipulating Majorana fermions for quantum computation.  The more recent solid state proposals noted above involving semiconductor heterostructures\cite{Sau} and non-centrosymmetric superconductors\cite{SatoFujimoto,PatrickProposal} utilize clever ways of creating an environment similar to the surface of a topological insulator (\emph{i.e.}, eliminating a sort of fermion doubling problem\cite{Shinsei}) in order to generate topological phases supporting Majorana modes.  

\begin{figure}
\centering{
  \includegraphics[width=3.2in]{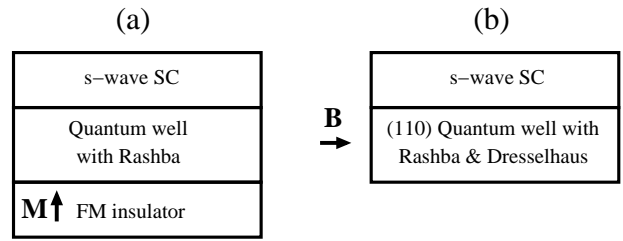}
  \caption{(a) Setup proposed by Sau \emph{et al}.\cite{Sau} for realizing a topological superconducting phase supporting Majorana fermions in a semiconductor quantum well with Rashba spin-orbit coupling.  The $s$-wave superconductor generates the pairing field in the well via the proximity effect, while the ferromagnetic insulator induces the Zeeman field required to drive the topological phase.  As noted by Sau \emph{et al.}, the Zeeman field can alternatively be generated by employing a magnetic semiconductor quantum well.  (b) Alternative setup proposed here.  We show that a (110)-grown quantum well with both Rashba and Dresselhaus spin-orbit coupling can be driven into a topological superconducting state by applying an in-plane magnetic field.  The advantages of this setup are that the Zeeman field is tunable, orbital effects are expected to be minimal, and the device is simpler, requiring neither a good interface with a ferromagnetic insulator nor the presence of magnetic impurities which provide an additional disorder source. }
  \label{QuantumWellSetups}}
\end{figure}

The present work is inspired by the semiconductor proposal of Sau \emph{et al}.\ in Ref.\ \onlinecite{Sau}, so we briefly elaborate on it here.  These authors demonstrated that a semiconductor with Rashba spin-orbit coupling, sandwiched between an $s$-wave superconductor and a ferromagnetic insulator as in Fig.\ \ref{QuantumWellSetups}(a), can realize a topological superconducting phase supporting Majorana modes.  The basic principle here is that the ferromagnetic insulator produces a Zeeman field perpendicular to the semiconductor, which separates the two spin-orbit-split bands by a finite gap.  If the Fermi level lies inside of this gap, a weak superconducting pair field generated via the proximity effect drives the semiconductor into a topological superconducting state that smoothly connects to a spinless $p_x+ip_y$ superconductor.  Sau \emph{et al}.\ also discussed how such a device can be exploited along the lines of the Fu-Kane proposal for topological quantum computation.  The remarkable aspect of this proposal is the conventional ingredients it employs---semiconductors benefit from many more decades of study compared to the relatively nascent topological insulators---making this a promising experimental direction.

The main question addressed in this paper is largely a practical one---can this proposed setup be further simplified and made more tunable, thus (hopefully) streamlining the route towards experimental realization of a topological superconducting phase in semiconductor devices?  To this end, there are two obvious modifications that one might try.  First, replacing the ferromagnetic insulator with an external magnetic field applied perpendicular to the semiconductor certainly simplifies the setup, but unfortunately induces undesirable orbital effects which change the problem significantly and likely spoil the topological phase.  The second obvious modification, then, would be applying an in-plane magnetic field.  While this sidesteps the problem of unwanted orbital effects, unfortunately in-plane fields do not open a gap between the spin-orbit-split bands in a Rashba-coupled semiconductor.  Physically, opening a gap requires a component of the Zeeman field perpendicular to the plane in which the electron spins orient; with Rashba coupling this always coincides with the semiconductor plane.  (See Sec.\ III for a more in-depth discussion.)  
 
Our main result is that a topological superconducting state supporting Majorana fermions \emph{can} be generated by in-plane magnetic fields if one alternatively considers a semiconductor grown along the (110) direction with both Rashba \emph{and} Dresselhaus coupling [see Fig.\ \ref{QuantumWellSetups}(b)].  What makes this possible in (110) semiconductors is the form of Dresselhaus coupling specific to this growth direction, which favors aligning the spins normal to the semiconductor plane.    
When Rashba coupling is also present, the two spin-orbit terms conspire to rotate the plane in which the spins orient away from the semiconductor plane.  In-plane magnetic fields then \emph{do} open a finite gap between the bands.  Under realistic conditions which we detail below, the proximity effect can then drive the system into a topological superconducting phase supporting Majorana modes, just as in the proposal from Ref.\ \onlinecite{Sau}.  

This alternative setup offers a number of practical advantages.  It eliminates the need for a good interface between the ferromagnetic insulator (or magnetic impurities intrinsic to the semiconductor\cite{Sau}), reducing considerably the experimental challenge of fabricating the device, while still largely eliminating undesired orbital effects.  Furthermore, explicitly controlling the Zeeman field in the semiconductor is clearly advantageous, enabling one to readily sweep across a quantum phase transition into the topological superconducting state and thus unambiguously identify the topological phase experimentally.  We propose that InSb quantum wells, which enjoy sizable Dresselhaus coupling and a large $g$-factor, may provide an ideal candidate for the semiconductor in such a device.  While not without experimental challenges (discussed in some detail below), we contend that this setup provides perhaps the simplest, most tunable semiconductor realization of a topological superconducting phase, so we hope that it will be pursued experimentally.  

The rest of the paper is organized as follows.  In Sec.\ II we provide a pedagogical overview of the proposal from Ref.\ \onlinecite{Sau}, highlighting the connection to a spinless $p_x+ip_y$ superconductor, which makes the existence of Majorana modes in this setup more intuitively apparent.  We also discuss in some detail the stability of the topological superconducting phase as well as several experimental considerations.  In Sec.\ III we introduce our proposal for (110) semiconductor quantum wells.  We show that the (110) quantum well Hamiltonian maps onto the Rashba-only model considered by Sau \emph{et al}.\ in an (unphysical) limit, and explore the stability of the topological superconductor here in the realistic parameter regime. Experimental issues related to this proposal are also addressed.  Finally, we summarize the results and discuss several future directions in Sec.\ IV.

\section{Overview of Sau-Lutchyn-Tewari-Das Sarma proposal}

To set the stage for our proposal, we begin by pedagogically reviewing the recent idea by Sau \emph{et al}.\ for creating Majorana fermions in a ferromagnetic insulator/semiconductor/s-wave superconductor hybrid system \cite{Sau} [see Fig.\ \ref{QuantumWellSetups}(a)].  These authors originally proved the existence of Majorana modes in this setup by explicitly solving the Bogoliubov-de Gennes Hamiltonian with a vortex in the superconducting order parameter.  An index theorem supporting this result was subsequently proven\cite{index}.  We will alternatively follow the approach employed in Ref.\ \onlinecite{FujimotoTSC} (see also Ref.\ \onlinecite{TopologicalSF}), and highlight the connection between the semiconductor Hamiltonian (in a certain limit) and a spinless $p_x+ip_y$ superconductor.  The advantage of this perspective is that the topological character of the proximity-induced superconducting state of interest becomes immediately apparent, along with the existence of a Majorana bound state at vortex cores.  In this way, one circumvents the cumbersome problem of solving the Bogoliubov-de Gennes equation for these modes.  The stability of the superconducting phase, which we will also discuss in some detail below, becomes more intuitive from this viewpoint as well.

\subsection{Connection to a spinless $p_x + ip_y$ superconductor}

Consider first an isolated zincblende semiconductor quantum well, grown along the (100) direction for concreteness.  Assuming layer (but not bulk) inversion asymmetry and retaining terms up to quadratic order in momentum\cite{SpinOrbitHigherOrder}, the relevant Hamiltonian reads
\begin{equation}
  H_0 = \int d^2{\bf r}\psi^\dagger \left[ -\frac{\nabla^2}{2m}-\mu - i\alpha (\sigma^x \partial_y-\sigma^y \partial_x) \right] \psi,
  \label{H0}
\end{equation}
where $m$ is the effective mass, $\mu$ is the chemical potential, $\alpha$ is the Rashba spin-orbit\cite{Rashba} coupling strength, and $\sigma^j$ are Pauli matrices that act on the spin degree of freedom in $\psi$.  (We set $\hbar = 1$ throughout.)  The Rashba terms above can be viewed as an effective magnetic field that aligns the spins in the quantum well plane, normal to their momentum.  Equation (\ref{H0}) admits two spin-orbit-split bands that appear `Dirac-like' at sufficiently small momenta where the $\nabla^2/2m$ kinetic term can be neglected.  The emergence of Majorana modes can ultimately be traced to this simple fact.

Coupling the semiconductor to a ferromagnetic insulator whose magnetization points perpendicular to the 2D layer is assumed to induce a Zeeman interaction 
\begin{equation}
  H_Z = \int d^2{\bf r} \psi^\dagger[V_z \sigma^z]\psi
  \label{HZ}
\end{equation}
but negligible orbital coupling.  Orbital effects will presumably be unimportant in the case where, for instance, $V_z$ arises primarily from exchange interactions rather than direct coupling of the spins to the field emanating from the ferromagnetic moments.  With this coupling, the spin-orbit-split bands no longer cross, and resemble a gapped Dirac point at small momenta.  Crucially, when $|\mu | < |V_z|$ the electrons in the quantum well then occupy only the lower band and exhibit a single Fermi surface.  We focus on this regime for the remainder of this section.  

What differentiates the present problem from a conventional single band (without spin-orbit coupling) is the structure of the wavefunctions inherited from the Dirac-like physics encoded in $H_0$ at small momenta.  To see this, it is illuminating to first diagonalize $H_0 + H_Z$ by writing
\begin{equation}
  \psi({\bf k}) = \phi_-({\bf k}) \psi_-({\bf k}) + \phi_+({\bf k})\psi_+({\bf k}),
\end{equation}
where $\psi_\pm$ annihilate states in the upper/lower bands and $\phi_\pm$ are the corresponding normalized wavefunctions,
\begin{eqnarray}
   \phi_+({\bf k}) &=& \left( \begin{array}{l}
         A_\uparrow(k) \\
        A_\downarrow(k) \frac{i k_x-k_y}{k} \end{array} \right) 
        \label{phiplus} \\
   \phi_-({\bf k}) &=& \left( \begin{array}{l}
         B_\uparrow(k)\frac{i k_x+k_y}{k} \\
        B_\downarrow(k) \end{array} \right).
        \label{phiminus}      
\end{eqnarray}
The expressions for $A_{\uparrow,\downarrow}$ and $B_{\uparrow,\downarrow}$ are not particularly enlightening, but for later we note the following useful combinations:
\begin{eqnarray}
  f_p(k) &\equiv& A_\uparrow A_\downarrow = B_\uparrow B_\downarrow = \frac{-\alpha k}{2\sqrt{V_z^2 + \alpha^2 k^2}}
  \\
  f_s(k) &\equiv& A_\uparrow B_\downarrow -B_\uparrow A_\downarrow = \frac{V_z}{\sqrt{V_z^2 + \alpha^2 k^2}}.
\end{eqnarray}
In terms of $\psi_\pm$, the Hamiltonian becomes
\begin{equation}
  H_0 + H_Z = \int d^2{\bf k}[\epsilon_+(k)\psi^\dagger_+({\bf k})\psi_+({\bf k}) + \epsilon_-(k)\psi_-^\dagger({\bf k})\psi_-({\bf k})],
\end{equation}
with energies
\begin{eqnarray}
  \epsilon_\pm(k) = \frac{k^2}{2m}-\mu \pm \sqrt{V_z^2 + \alpha^2 k^2}.
\end{eqnarray}

Now, when the semiconductor additionally comes into contact with an s-wave superconductor, a pairing term will be generated via the proximity effect, so that the full Hamiltonian describing the quantum well becomes 
\begin{equation}
  H = H_0 + H_Z + H_{SC}
\end{equation} with
\begin{equation}
  H_{SC} = \int d^2{\bf r}[\Delta \psi^\dagger_\uparrow \psi^\dagger_\downarrow + h.c.].
  \label{Hsc}
\end{equation}
(We note that $H$ is a continuum version of the lattice model discussed in Ref.\ \onlinecite{FujimotoColdAtoms} in the context of topological superfluids of cold fermionic atoms.)
Rewriting $H_{SC}$ in terms of $\psi_\pm$ and using the wavefunctions in Eqs.\ (\ref{phiplus}) and (\ref{phiminus}) yields
\begin{eqnarray}
  H_{SC} &=& \int d^2{\bf k}\bigg{[}\Delta_{+-}(k) \psi^\dagger_+({\bf k})\psi^\dagger_-(-{\bf k}) 
  \nonumber \\
  &+& \Delta_{--}({\bf k})\psi^\dagger_-({\bf k})\psi^\dagger_-(-{\bf k}) 
  \nonumber \\
  &+& \Delta_{++}({\bf k})\psi^\dagger_+({\bf k})\psi^\dagger_+(-{\bf k}) + h.c. \bigg{]},
  \label{Hsc2}
\end{eqnarray}
with
\begin{eqnarray}
  \Delta_{+-}(k) &=& f_s(k) \Delta
  \\
  \Delta_{++}({\bf k}) &=& f_p(k)\left(\frac{k_y + i k_x}{k}\right) \Delta
  \\   
  \Delta_{--}({\bf k}) &=& f_p(k)\left(\frac{k_y - i k_x}{k}\right) \Delta
.
\end{eqnarray}
The proximity effect thus generates not only interband s-wave pairing encoded in the first term, but also \emph{intra}band $p_x\pm i p_y$ pairing with opposite chirality for the upper/lower bands.  This is exactly analogous to spin-orbit-coupled superconductors, where the pairing consists of spin-singlet and spin-triplet components due to non-conservation of spin\cite{SpinOrbitSC}.  

We can now immediately understand the appearance of a topological superconducting phase in this system.  Consider $\Delta$ much smaller than the spacing $|V_z-\mu|$ to the upper band.  In this case the upper band plays essentially no role and can simply be projected away by sending $\psi_+\rightarrow 0$ above.  The problem then maps onto that of spinless fermions with $p_x+ip_y$ pairing, which is the canonical example of a topological superconductor supporting a single Majorana bound state at vortex cores\cite{ReadGreen,Ivanov}.  (The dispersion $\epsilon_-(k)$ is, however, somewhat unconventional.  But one can easily verify that the dispersion can be smoothly deformed into a conventional $k^2/2m-\mu$ form, with $\mu>0$, without closing a gap.)  Thus, in this limit introducing a vortex in the order parameter $\Delta$ must produce a single Majorana bound state in this semiconductor context as well.  

We emphasize that in the more general case where $\Delta$ is not negligible compared to $|V_z-\mu|$, the mapping to a spinless $p_x+ip_y$ superconductor is no longer legitimate.  
Nevertheless, since the presence of a Majorana fermion has a topological origin, it can not disappear as long as the bulk excitation gap remains finite.  We will make extensive use of this fact in the remainder of the paper.  Here we simply observe that the topological superconducting state and Majorana modes will persist even when one incorporates both bands---which we do hereafter---provided the pairing $\Delta$ is sufficiently small that the gap does not close, as found explicitly by studying the full unprojected Hamiltonian with a vortex in Ref.\ \onlinecite{Sau}.

It is also important to stress that when $\Delta$ greatly exceeds $V_z$, it is the Zeeman field that essentially plays no role.  A topological superconducting state is no longer expected in this limit, since one is not present when $V_z = 0$.  Thus as $\Delta$ increases, the system undergoes a quantum phase transition from a topological to an ordinary superconducting state, as discussed by Sau \emph{et al}.\cite{Sau} and Sato \emph{et al}.\cite{FujimotoColdAtoms} in the cold-atoms context. The transition is driven by the onset of interband $s$-wave pairing near zero momentum.

\subsection{Stability of the topological superconducting phase}

The stability of the topological superconducting state was briefly discussed in Ref.\ \onlinecite{Sau}, as well as Ref.\ \onlinecite{FujimotoColdAtoms} in the cold-atoms setting.  Here we address this issue in more detail, with the aim of providing further intuition as well as guidance for experiments.  
Given the competition between ordinary and topological superconducting order inherent in the problem, it is useful to explore, for instance, how the chemical potential, spin-orbit strength, proximity-induced pair field, and Zeeman field should be chosen so as to maximize the 
bulk excitation gap in the topological phase of interest.  Furthermore, what limits the size of this gap, and how does it decay as these parameters are tuned away from the point of maximum stability?  And how are other important factors such as the density impacted by the choice of these parameters?

Solving the full Bogoliubov-de Gennes Hamiltonian assuming uniform $\Delta$ yields energies that satisfy
\begin{eqnarray}
  E_\pm^2 &=& 4|\Delta_{++}|^2 +\Delta_{+-}^2 + \frac{\epsilon_+^2 + \epsilon_-^2}{2} 
  \nonumber \\
  &\pm& |\epsilon_+-\epsilon_-|\sqrt{\Delta_{+-}^2 + \frac{(\epsilon_+ + \epsilon_-)^2}{4}}.
\end{eqnarray}
We are interested in the lower branch $E_-(k)$, in particular its value at zero momentum and near the Fermi surface.  The minimum of these determines the bulk superconducting gap, $E_g \equiv \Delta G(\frac{\mu}{V_z},\frac{m\alpha^2}{V_z},\frac{\Delta}{V_z})$.  

To make the topological superconducting state as robust as possible, one clearly would like to maximize the $p$-wave pairing at the Fermi momentum,
\begin{equation} 
 k_F = \sqrt{2m\left[m\alpha^2 + \mu + \sqrt{V_z^2 + m\alpha^2(m\alpha^2 + 2 \mu)}\right]}.
 \label{kF}
\end{equation}
Doing so requires $m \alpha^2/V_z \gg 1$.  In this limit we have $|\Delta_{++}(k_F)| \sim \Delta/2$ while the $s$-wave pairing at the Fermi momentum is negligible, $\Delta_{+-}(k_F) \sim 0$.  We thus obtain
\begin{equation}
  E_-(k_F) \sim \Delta,
  \label{EkF}
\end{equation}  
which increases monotonically with $\Delta$.  At zero momentum, however, we have
\begin{equation}
  E_-(k = 0) = |V_z-\sqrt{\Delta^2+\mu^2}|.
  \label{Ek0}
\end{equation}
This initially \emph{decreases} with $\Delta$ as interband $s$-wave pairing begins to set in, and vanishes when $\Delta = \sqrt{V_z^2-\mu^2}$ signaling the destruction of the topological superconducting state\cite{Sau,FujimotoColdAtoms}.  It follows that for a given $V_z$, the topological superconductor is most robust when $m \alpha^2/V_z \gg 1$, $\mu = 0$, and $\Delta = V_z/2$; here the bulk excitation gap is maximized and given by $E_g = V_z/2$.  

\begin{figure}
\centering{
\subfigure{
\includegraphics[width=3.5in]{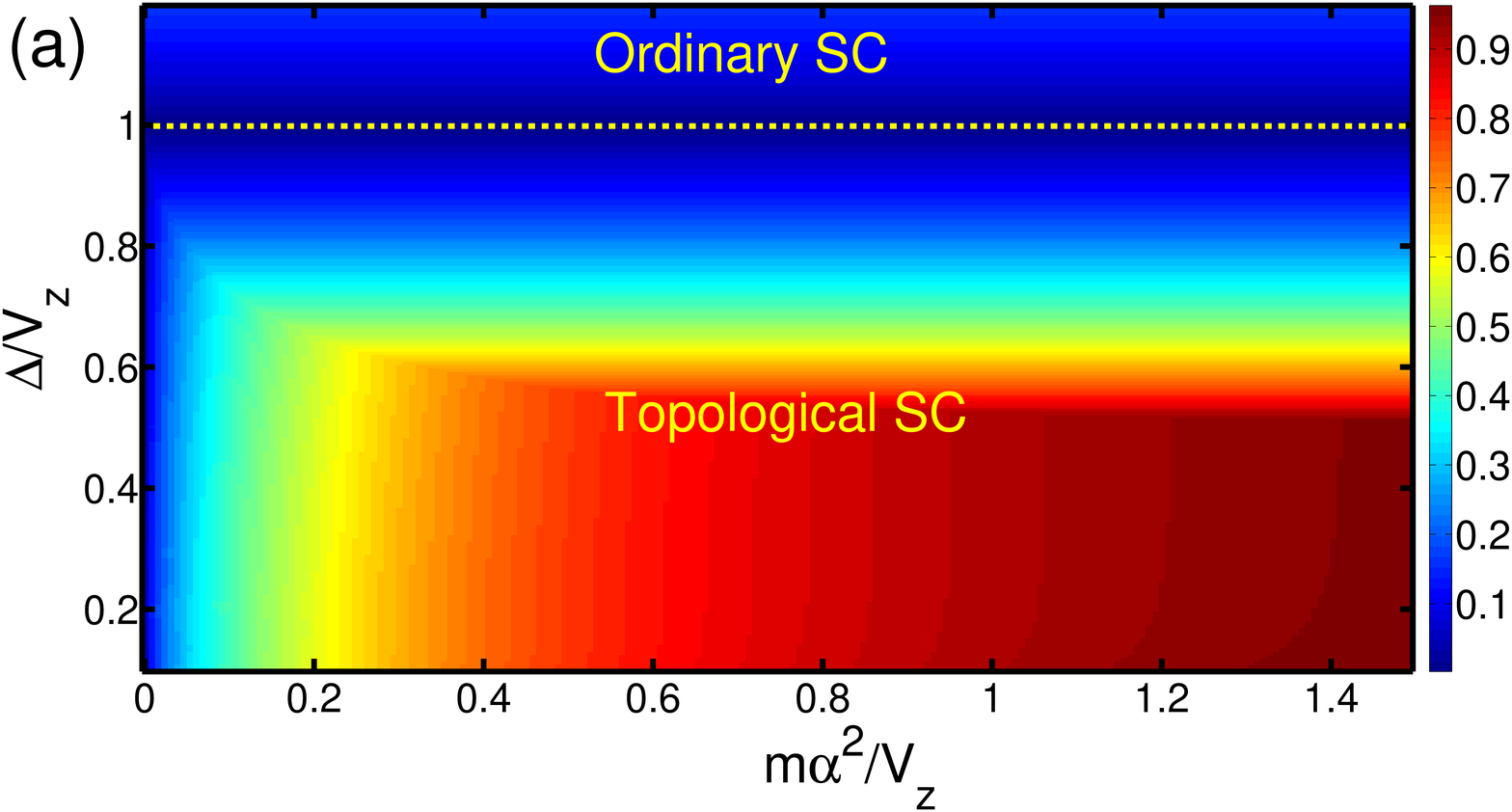}
\label{fig:subfig1}
}
\subfigure{
\includegraphics[width=3.5in]{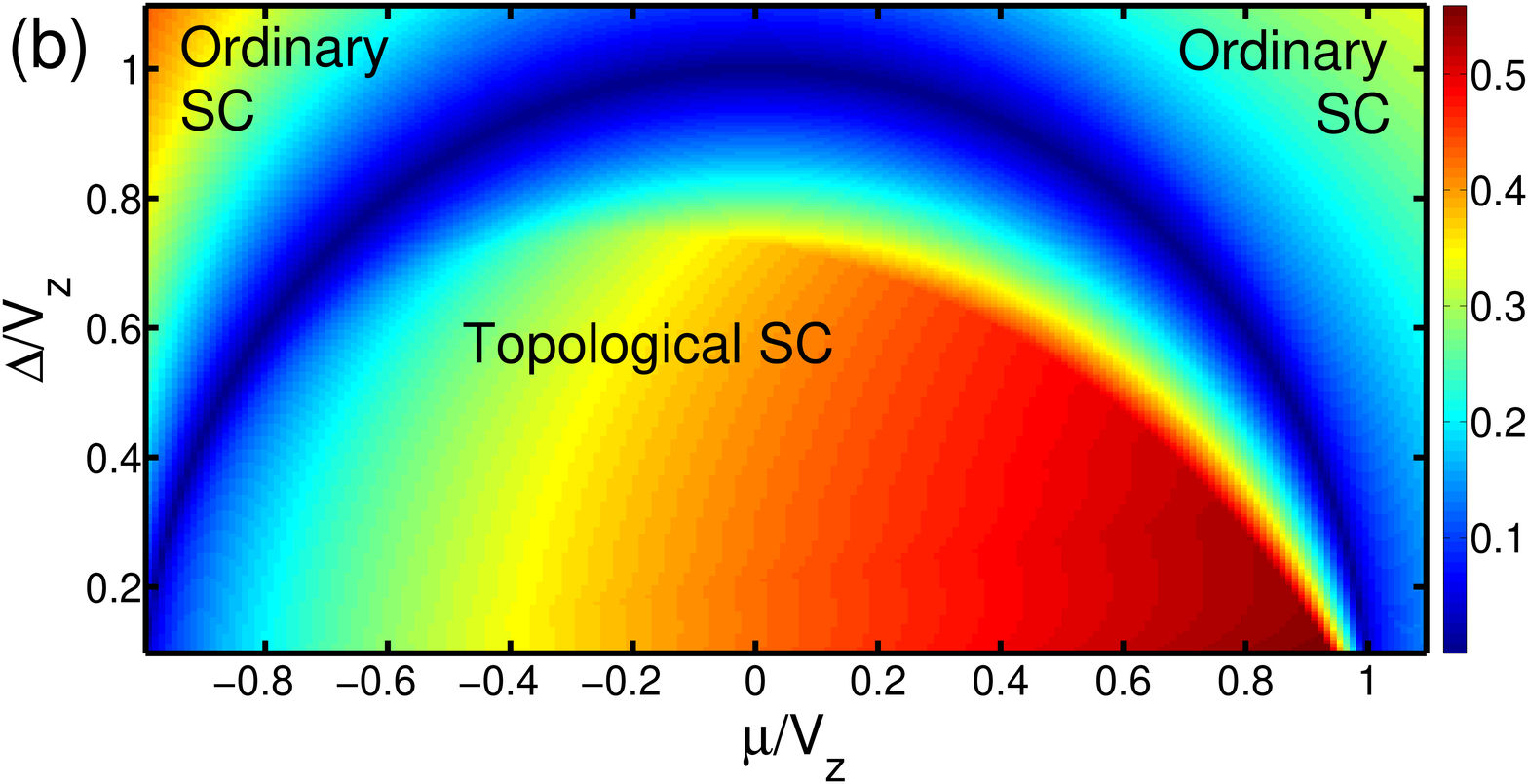}
\label{fig:subfig2}}
\caption{Excitation gap $E_g$ normalized by $\Delta$ in the proximity-induced superconducting state of a Rashba-coupled quantum well adjacent to a ferromagnetic insulator.  In (a), the chemical potential is chosen to be $\mu = 0$.  For $\Delta/V_z <1$ the system realizes a topological superconducting phase supporting a single Majorana mode at a vortex core, while for $\Delta/V_z >1$ an ordinary superconducting state emerges.  In the topological phase, the gap is maximized when $\Delta/V_z = 1/2$ and $m\alpha^2/V_z \gg 1$, where it is given by $E_g = V_z/2$.  In contrast, the gap vanishes as $m\alpha^2/V_z\rightarrow 0$ because the effective $p$-wave pair field at the Fermi momentum vanishes in this limit.  In (b), we have taken $m\alpha^2/V_z = 0.1$ to illustrate that $V_z$ can exceed $m\alpha^2$ by more than an order of magnitude and still yield a sizable gap in the topological superconducting phase.  } 
\label{GapFig}}
\end{figure}

As will become clear below, for practical purposes it is also useful to explore the limit where $V_z$ is much \emph{larger} than both $\Delta$ and $m \alpha^2$.  Here the gap is determined solely by the $p$-wave pair field near the Fermi surface [except for $\mu$ very close to $V_z$, where it follows from Eq.\ (\ref{Ek0})].  This pairing will certainly be reduced compared to the $m \alpha^2/V_z \gg 1$ limit, because the lower band behaves like a conventional quadratically dispersing band in the limit $m\alpha^2/V_z \rightarrow 0$.  To leading order in $m\alpha^2/V_z$ and $\Delta/V_z$, the gap is given by
\begin{equation}
  E_g \approx \sqrt{\frac{2m\alpha^2}{V_z}\left(1+\frac{\mu}{V_z}\right)} \Delta.
\end{equation}
There are two noteworthy features of this expression.  First, although the gap indeed vanishes as $m\alpha^2/V_z \rightarrow 0$, it does so very slowly; $V_z$ can exceed $m \alpha^2$ by more than an order of magnitude and still yield a gap that is a sizable fraction of the bare proximity-induced $\Delta$.  Second, in this limit the gap can be enhanced by raising $\mu$ near the bottom of the upper band.  

These results are graphically summarized in Fig.\ \ref{GapFig}, which displays the gap $E_g$ normalized by $\Delta$.  Figure \ref{GapFig}(a) assumes $\mu = 0$ and illustrates the dependence on $m\alpha^2/V_z$ and $\Delta/V_z$; Fig.\ \ref{GapFig}(b) assumes $m\alpha^2/V_z = 0.1$ and illustrates the dependence on $\Delta/V_z$ and $\mu/V_z$.  Note that despite the relatively small value of $m \alpha^2/V_z$ chosen here, the gap remains a sizable fraction of $\Delta$ over much of the topological superconductor regime.

\subsection{Experimental considerations}

The quantity $m\alpha^2$ comprises a crucial energy scale regarding experimental design.  Ideally, this should be as large as possible for at least two reasons.  First, the scale of $m\alpha^2$ limits how large a Zeeman splitting $V_z$ is desirable.  If $m\alpha^2/V_z$ becomes too small, then as discussed above the effective $p$-wave pairing at the Fermi surface will eventually be strongly suppressed compared to $\Delta$, along with the bulk excitation gap.  At the same time, having a large $V_z$ is advantageous in that the topological superconductor can then exist over a broad range of densities.  This leads us to the second reason why large $m\alpha^2$ is desired: this quantity strongly impacts the density in the topological superconductor regime,
\begin{eqnarray}
  n = \frac{(m\alpha)^2}{2\pi}\left[1+\frac{\mu}{m\alpha^2}+\sqrt{1+\left(\frac{V_z}{m\alpha^2}\right)^2+\frac{2\mu}{m\alpha^2}}\right]. 
  \label{n} 
\end{eqnarray}
One should keep in mind that if the density is too small, disorder may dominate the physics\cite{LowDensity2DEG}.\footnote{Lowering the density does, however, lead to a smaller Fermi energy and thus a larger `mini-gap' associated with the vortex-core bound states.  Large mini-gaps are ultimately desirable for topological quantum computation.  In this paper we are concerned with a more modest issue---namely, simply finding a stable topological superconducting phase in the first place.  With this more limited goal in mind, large densities are clearly desired to reduce disorder effects.  }  

Experimental values for the Rashba coupling $\alpha$ depend strongly on the properties of the quantum well under consideration, and, importantly, are tunable in gated systems\cite{RashbaTuning} (see also Ref.\ \onlinecite{RashbaTuningB}).  In GaAs quantum wells, for instance, $\alpha \approx  0.005$eV\AA\cite{GaAsSpinOrbit1} and $\alpha \approx 0.0015$eV\AA\cite{GaAsSpinOrbit2} have been measured.  Using the effective mass $m = 0.067m_e$ ($m_e$ is the bare electron mass), these correspond to very small energy scales $m\alpha^2 \sim 3$mK for the former and a scale an order of magnitude smaller for the latter.  In the limit $m\alpha^2/V_z \ll 1$, Eq.\ (\ref{n}) yields a density for the topological superconductor regime of $n \sim 10^7$cm$^{-2}$ and $\sim 10^6$cm$^{-2}$, respectively.  Disorder likely dominates at such low densities.  Employing Zeeman fields $V_z$ which are much larger than $m\alpha^2$ can enhance these densities by one or two orders of magnitude without too dramatically reducing the gap (the density increases much faster with $V_z$ than the gap decreases), though this may still be insufficient to overcome disorder effects.  

Due to their stronger spin-orbit coupling, quantum wells featuring heavier elements such as In and Sb appear more promising.  A substantially larger $\alpha \approx 0.06$eV\AA~ has been measured\cite{InAsSpinOrbit} in InAs quantum wells with effective mass $m \approx 0.04m_e$, yielding a much greater energy scale $m \alpha^2 \sim 0.2$K.  The corresponding density in the $m\alpha^2/V_z \ll 1$ limit is now $n \sim 10^8$cm$^{-2}$.  While still small, a large Zeeman field corresponding to $m\alpha^2/V_z = 0.01$ raises the density to a more reasonable value of $n \sim 10^{10}$cm$^{-2}$.  
As another example, the Rashba coupling in InGaAs quantum wells with $m \approx 0.05m_e$ was tuned over the range $\alpha \sim 0.05-0.1$eV\AA~ with a gate\cite{RashbaTuning}, resulting in a range of energy scales $m\alpha^2 \sim 0.2-0.8$K.  The densities here are even more promising, with $n \sim 10^8-10^9$cm$^{-2}$ in the limit $m\alpha^2/V_z \ll 1$; again, these can be enhanced significantly by considering $V_z$ large compared to $m\alpha^2$.  

To conclude this section, we comment briefly on the setups proposed by Sau \emph{et al}., wherein the Zeeman field arises either from a proximate ferromagnetic insulator or magnetic impurities in the semiconductor.  In principle, the Rashba coupling and chemical potential should be separately tunable in either case by applying a gate voltage and adjusting the Fermi level in the $s$-wave superconductor.  The strength of the Zeeman field, however, will largely be dictated by the choice of materials, doping, geometry, \emph{etc}.  Unless the value of $m\alpha^2$ can be greatly enhanced compared to the values quoted above, it may be advantageous to consider Zeeman fields which are much larger than this energy scale, in order to raise the density at the expense of suppressing the bulk excitation gap somewhat.  A good interface between the ferromagnetic insulator and the quantum well will be necessary to achieve a large $V_z$, if this setup is chosen.  Allowing $V_z$ to arise from magnetic impurities eliminates this engineering challenge, but has the drawback that the dopants provide another disorder source which can deleteriously affect the device's mobility\cite{MagneticSCdisorder}.  Nevertheless, since semiconductor technology is so well advanced, it is certainly worth pursuing topological phases in this setting, especially if alternative setups minimizing these challenges can be found.  Providing one such alternative is the goal of the next section.  

\section{Proposed setup for (110) quantum wells}

We now ask whether one can make the setup proposed by Sau \emph{et al}.\ simpler and more tunable by replacing the ferromagnetic insulator (or magnetic impurities embedded in the semiconductor) responsible for the Zeeman field with an experimentally controllable parameter.  As mentioned in the introduction, the most naive possible way to achieve this would be to do away with the magnetic insulator (or magnetic impurities) and instead simply apply an external magnetic field perpendicular to the semiconductor.  In fact, this possibility was pursued earlier in Refs.\ \onlinecite{FujimotoTSC} and \onlinecite{SatoFujimoto}.  It is far from obvious, however, that the Zeeman field dominates over orbital effects here, which was a key ingredient in the proposal by Sau \emph{et al}.  Thus, these references focused on the regime where the Zeeman field was smaller than $\Delta$, which is insufficient to drive the topological superconducting phase.  (We note, however, that a proximity-induced spin-triplet order parameter, if large enough, was found to stabilize a topological state\cite{SatoFujimoto}.)  An obvious alternative would be applying a parallel magnetic field, along the quantum well plane, since this (largely) rids of the unwanted orbital effects.  This too is insufficient, since replacing $V_z\sigma^z$ with $V_y \sigma^y$ in Eq.\ (\ref{HZ}) does not gap out the bands at $k = 0$, but only shifts the crossing to finite momentum.  

\subsection{Topological superconducting phase in a (110) quantum well}

We will show that if one alternatively considers a zincblende quantum well grown along the (110) direction, a topological superconducting state \emph{can} be driven by application of a parallel magnetic field.  What makes this possible in (110) quantum wells is their different symmetry compared to (100) quantum wells.  Assuming layer inversion symmetry is preserved, the most general Hamiltonian for the well up to quadratic order in momentum\cite{SpinOrbitHigherOrder} is
\begin{equation}
  \mathcal{H}_0 = \int d^2{\bf r}\psi^\dagger\left[-\left(\frac{\partial_x^2}{2m_x} + \frac{\partial_y^2}{2m_y}\right)-\mu -i \beta \partial_x \sigma^z\right]\psi
\end{equation}
Here we allow for anisotropic effective masses $m_{x,y}$ due to a lack of in-plane rotation symmetry, and $\beta$ is the Dresselhaus spin-orbit\cite{Dresselhaus} coupling strength.  Crucially, the Dresselhaus term favors alignment of the spins \emph{normal} to the plane, in contrast to the Rashba coupling in Eq.\ (\ref{H0}) which aligns spins \emph{within} the plane.  Although we did not incorporate Dresselhaus terms in the previous section, we note that in a (100) quantum well they, too, favor alignment of spins within the plane.

As an aside, we note that the above Hamiltonian has been of interest in the spintronics community because it preserves the $S^z$ component of spin as a good quantum number, resulting in long lifetimes for spins aligned normal to the quantum well\cite{LongSpinLifetime}.  ($\mathcal{H}_0$ also exhibits a `hidden' SU(2) symmetry\cite{PersistentSpinHelix} which furthered interest in this model, but this is not a microscopic symmetry and will play no role here.)  We are uninterested in spin lifetimes, however, and wish to explicitly break layer inversion symmetry by imbalancing the quantum well using a gate voltage and/or chemical means.  The Hamiltonian for the (110) quantum well then becomes $\mathcal{H}^{(110)} = \mathcal{H}_0 + \mathcal{H}_R$, where 
\begin{equation}
  \mathcal{H}_R =  \int d^2{\bf r}\psi^\dagger\left[-i(\alpha_x \sigma^x \partial_y-\alpha_y\sigma^y\partial_x\right)]\psi
\end{equation}
represents the induced Rashba spin-orbit coupling terms up to linear order in momentum.
While one would naively expect $\alpha_x = \alpha_y$ here, band structure effects will generically lead to unequal coefficients, again due to lack of rotation symmetry.  We can recast the quantum well Hamiltonian into a more useful form by rescaling coordinates so that $\partial_x \rightarrow (m_x/m_y)^{1/4}\partial_x$ and $\partial_y \rightarrow (m_y/m_x)^{1/4}\partial_y$.  We then obtain
\begin{eqnarray}
  \mathcal{H}^{(110)} &=& \int d^2{\bf r} \psi^\dagger \bigg{[}-\frac{\nabla^2}{2m_*}-\mu-i\lambda_D \partial_x\sigma^z
  \nonumber \\
  &-& i\lambda_R(\sigma^x \partial_y-\gamma \sigma^y \partial_x)\bigg{]}\psi.
  \label{H110}
\end{eqnarray}  
The effective mass is $m_* = \sqrt{m_x m_y}$ and the spin-orbit parameters are $\lambda_D = \beta (m_x/m_y)^{1/4}$, $\lambda_R = \alpha_x(m_y/m_x)^{1/4}$, and $\gamma = (\alpha_y/\alpha_x)\sqrt{m_x/m_y}$.

With both Dresselhaus and Rashba terms present, the spins will no longer align normal to the quantum well, but rather lie within the plane perpendicular to the vector $\lambda_D{\hat{\bf y}} + \gamma \lambda_R{\hat{\bf z}}$.    
Consider for the moment the important special case $\gamma = 0$ and $\lambda_D = \lambda_R$.  In this limit, $\mathcal{H}^{(110)}$ becomes essentially \emph{identical} to Eq.\ (\ref{H0}), with the important difference that here the spins point in the $(x,z)$ plane rather than the $(x,y)$ plane.  It follows that a field applied along the $y$ direction,
\begin{equation}
  \mathcal{H}_Z = \int d^2{\bf r} \psi^\dagger[V_y \sigma^y]\psi,
\end{equation}
with $V_y = g\mu_B B_y/2$, then plays \emph{exactly} the same role as the Zeeman term $V_z$ in Sau \emph{et al}.'s proposal\cite{Sau} discussed in the preceding section---the bands no longer cross at zero momentum, and only the lower band is occupied when $|\mu| < |V_y|$.  In this regime, when the system comes into contact with an $s$-wave superconductor, the proximity effect generates a topological superconducting state supporting Majorana fermions at vortex cores, provided the induced pairing in the well is not too large\cite{Sau}.  

The full problem we wish to study, then, corresponds to a (110) quantum well with both Dresselhaus and Rashba coupling, subjected to a parallel magnetic field and contacted to an $s$-wave superconductor.  The complete Hamiltonian is 
\begin{equation}
  \mathcal{H} = \mathcal{H}^{(110)} + \mathcal{H}_Z + \mathcal{H}_{SC},
\end{equation}
with $\mathcal{H}_{SC}$ the same as in Eq.\ (\ref{Hsc}).  Of course in a real system $\gamma$ will be non-zero, and likely of order unity, and $\lambda_R$ generally differs from $\lambda_D$.  The question we must answer then is how far the topological superconducting phase survives as we increase $\gamma$ from zero and change the ratio $\lambda_R/\lambda_D$ from unity.  Certainly our proposal will be viable only if this state survives relatively large changes in these parameters.

\subsection{Stability of the topological superconducting phase in (110) quantum wells}

To begin addressing this issue, it is useful to proceed as in the previous section and express the Hamiltonian in terms of operators $\psi_\pm^\dagger({\bf k})$ which add electrons to the upper/lower bands:
\begin{eqnarray}
  \mathcal{H} &=& \int d^2{\bf k}[\tilde \epsilon_+({\bf k})\psi^\dagger_+({\bf k})\psi_+({\bf k}) + \tilde \epsilon_-({\bf k})\psi_-^\dagger({\bf k})\psi_-({\bf k})]
  \nonumber \\
  &+&\bigg{[}\tilde\Delta_{+-}({\bf k}) \psi^\dagger_+({\bf k})\psi^\dagger_-(-{\bf k}) 
  + \tilde\Delta_{--}({\bf k})\psi^\dagger_-({\bf k})\psi^\dagger_-(-{\bf k}) 
  \nonumber \\
  &+& \tilde\Delta_{++}({\bf k})\psi^\dagger_+({\bf k})\psi^\dagger_+(-{\bf k}) + h.c. \bigg{]}.
\end{eqnarray}
The energies $\tilde \epsilon_\pm$ are given by
\begin{eqnarray}
  \tilde \epsilon_\pm({\bf k}) &=& \frac{k^2}{2m}-\mu  \pm \delta \tilde\epsilon({\bf k})
  \nonumber \\ 
  \delta \tilde \epsilon({\bf k}) &=& \sqrt{(V_y-\gamma \lambda_R k_x)^2 + (\lambda_D k_x)^2 + (\lambda_R k_y)^2},
  \label{epsilontilde}
\end{eqnarray}
while the interband $s$- and intraband $p$-wave pair fields now satisfy
\begin{eqnarray}
  |\tilde \Delta_{+-}({\bf k})|^2 &=& \frac{\Delta^2}{2}\bigg{[}1-\frac{(\lambda_D^2 + \gamma^2\lambda_R^2)k_x^2 + \lambda_R^2 k_y^2-V_y^2}{\delta\tilde\epsilon({\bf k})\delta \tilde \epsilon(-{\bf k})}\bigg{]}
  \nonumber \\
  |\tilde \Delta_{++}({\bf k})|^2 &=& |\tilde \Delta_{--}({\bf k})|^2 
  \label{Deltatilde} \\
  &=& \frac{\Delta^2}{8}\bigg{[}1+\frac{(\lambda_D^2 + \gamma^2\lambda_R^2)k_x^2 + \lambda_R^2 k_y^2-V_y^2}{\delta\tilde\epsilon({\bf k})\delta \tilde \epsilon(-{\bf k})}\bigg{]}.
  \nonumber
\end{eqnarray}

Increasing $\gamma$ from zero to of order unity affects the above pair fields rather weakly.  
The dominant effect of $\gamma$, which can be seen from Eq.\ (\ref{epsilontilde}), is to lift the $k_x\rightarrow -k_x$ symmetry of the $\Delta = 0$ bands.  Physically, this symmetry breaking arises because when $\gamma \neq 0$ the spins lie within a plane that is not perpendicular to the magnetic field.  This, in turn, suppresses superconductivity since states with ${\bf k}$ and $-{\bf k}$ will generally have different energy.  While in this case the Bogoliubov-de Gennes equation no longer admits a simple analytic solution, one can numerically compute the bulk energy gap for the uniform superconducting state, $\mathcal{E}_g \equiv \Delta \mathcal{G}(\frac{\mu}{V_y},\frac{m\lambda_D^2}{V_y},\frac{\Delta}{V_y},\frac{\lambda_R}{\lambda_D},\gamma)$, to explore the stability of the topological superconducting phase.

\begin{figure}
\centering
\subfigure{
\includegraphics[width=3.5in]{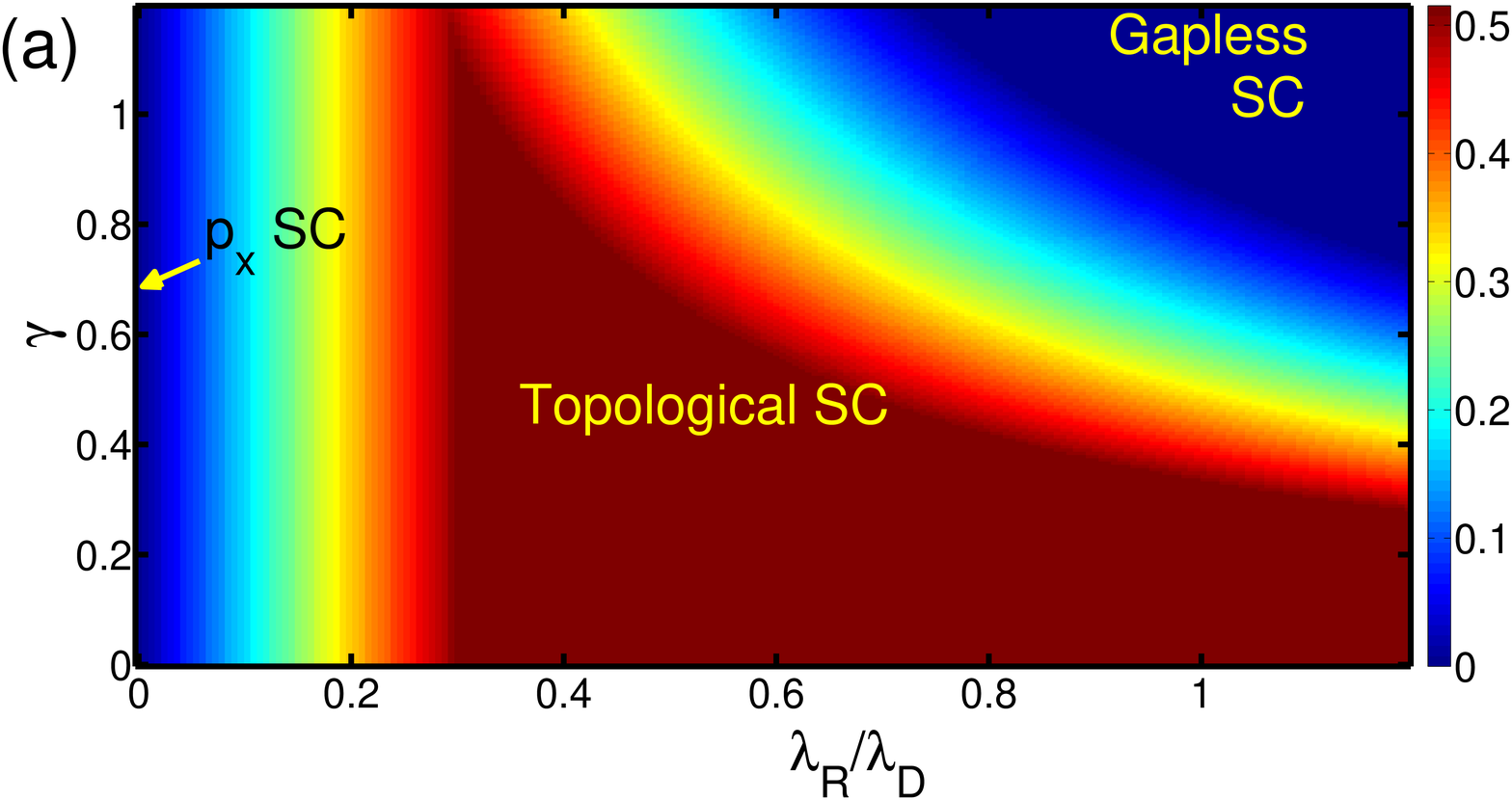}
\label{fig:110subfig1}
}
\subfigure{
\includegraphics[width=3.5in]{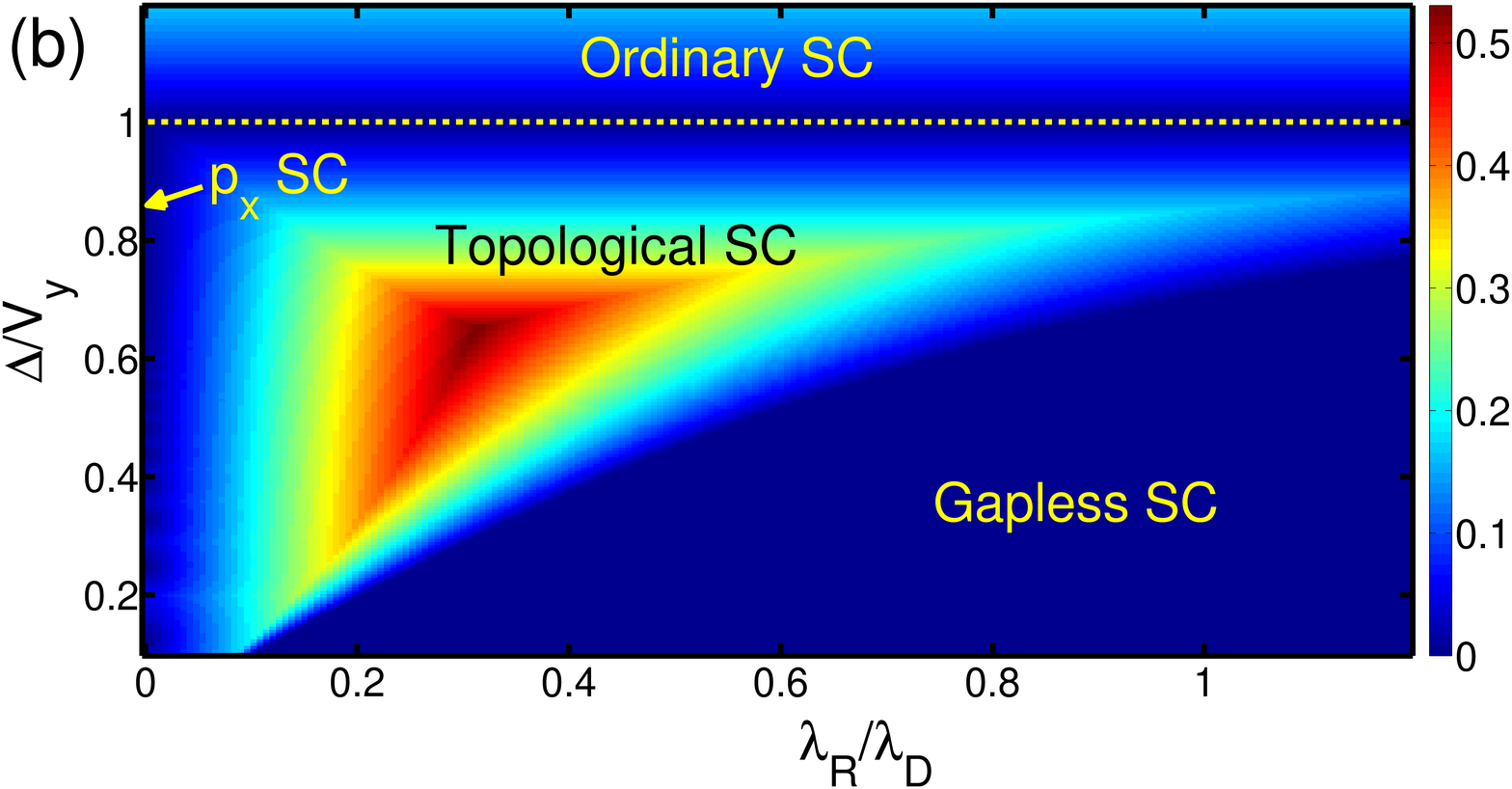}
\label{fig:110subfig2}
}
\subfigure{
\includegraphics[width=3.5in]{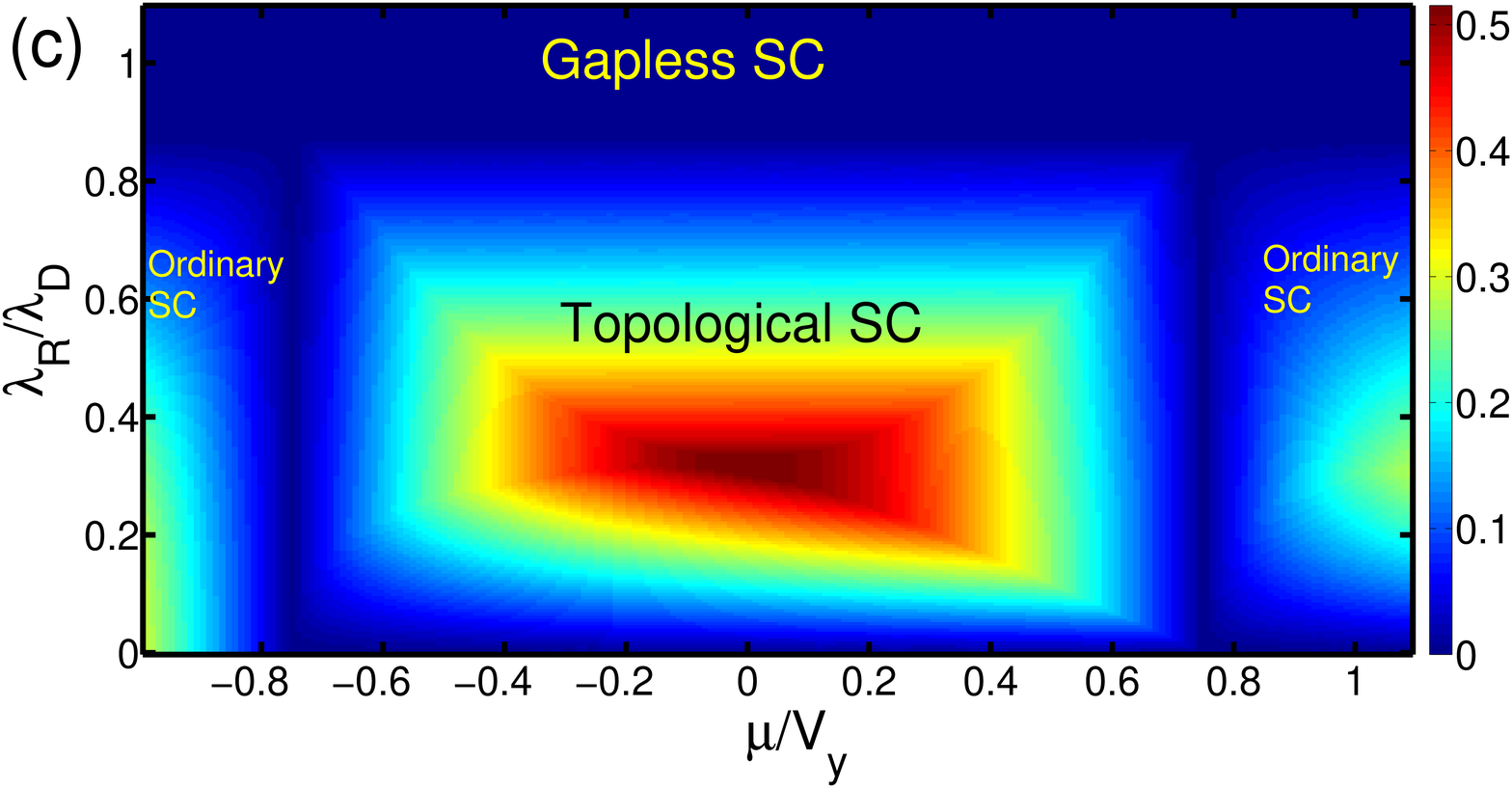}
\label{fig:110subfig3}
} \caption{Excitation gap $\mathcal{E}_g$ normalized by $\Delta$ in the proximity-induced superconducting state of a (110) quantum well, with both Rashba and Dresselhaus spin-orbit coupling, in a parallel magnetic field.  In (a) we set $\mu = 0$, $m\lambda_D^2/V_y = 2$, and $\Delta/V_y = 0.66$, and illustrate the dependence of the gap on the Rashba coupling anisotropy $\gamma$ as well as $\lambda_R/\lambda_D$.  When $\gamma = 0$ and $\lambda_R/\lambda_D = 1$, the problem maps onto the Rashba-only model considered by Sau \emph{et al}.\cite{Sau}  Remarkably, the gap survives unaltered here even in the physically relevant case with $\gamma$ of order one, provided the Rashba coupling is reduced. 
In (b) and (c), we focus on the realistic case with $\gamma = 1$ to illustrate the stability of the topological phase in more detail.  We take $\mu = 0$ and $m\lambda_D^2/V_y = 2$ in (b), and allow $\Delta/V_y$ as well as $\lambda_R/\lambda_D$ to vary.  In (c), we fix $\Delta/V_y = 0.66$ and $m\lambda_D^2/V_y = 2$, allowing $\mu/V_y$ and $\lambda_R/\lambda_D$ to vary.}
\label{GapFig110}
\end{figure}

Consider first the illustrative case with $\mu = 0$, $m \lambda_D^2/V_y = 2$, and $\Delta/V_y = 0.66$.  The corresponding gap as a function of $\lambda_R/\lambda_D$ and $\gamma$ appears in Fig.\ \ref{GapFig110}(a).  At $\lambda_R/\lambda_D=1$ and $\gamma = 0$, where our proposal maps onto that of Sau \emph{et al}., the gap is $E_g \approx 0.52 \Delta$, somewhat reduced from its maximum value since we have taken $\Delta/V_y > 1/2$.  Remarkably, as the figure demonstrates \emph{this gap persists unaltered even beyond $\gamma = 1$, provided the scale of Rashba coupling $\lambda_R/\lambda_D$ is suitably reduced}.  Throughout this region, the lowest-energy excitation is created at zero momentum, where the energy gap is simply $\mathcal{E}_g = V_y-\Delta$.  This clearly demonstrates the robustness of the topological superconducting state well away from the Rashba-only model considered by Sau \emph{et al}., and supports the feasibility of our modified proposal in (110) quantum wells.  

Let us understand the behavior of the gap displayed in Fig.\ \ref{GapFig110}(a) in more detail.  As described above, the plane in which the spins reside is tilted away from the $(x,z)$ plane by an angle $\theta = \cos^{-1}[1/\sqrt{1+(\gamma \lambda_R/\lambda_D)^2}]$.  Non-zero $\theta$ gives rise to the anisotropy under $k_x \rightarrow -k_x$, which again tends to suppress superconductivity.  One can see here that reducing $\lambda_R/\lambda_D$ therefore can compensate for an increase in $\gamma$, leading to the rather robust topological superconducting phase evident in the figure. 

On the other hand, at fixed $\lambda_R/\lambda_D$ which is sufficiently large ($\gtrsim 0.3$ in the figure), increasing $\gamma$ eventually results in the minimum energy excitation occurring at $k_y = 0$ and $k_x$ near the Fermi momentum.  Further increasing $\gamma$ then shrinks the gap and eventually opens pockets of gapless excitations, destroying the topological superconductor.  Conversely, if $\lambda_R/\lambda_D$ is sufficiently small ($\lesssim 1/3$ in the figure), the gap becomes independent of $\gamma$.  In this region the minimum energy excitations are created at $k_x = 0$ and $k_y$ near the Fermi momentum.  As $\lambda_R/\lambda_D\rightarrow 0$, the lower band transitions from a gapped topological $p_x+ip_y$ superconductor to a gapless nodal $p_x$ superconductor.  This follows from Eq.\ (\ref{Deltatilde}), which in the limit $\lambda_R = 0$ yields a pair field $\tilde\Delta_{--} = \Delta \lambda_D k_x/[2 \sqrt{V_y^2 + \lambda_D^2 k_x^2}]$ that vanishes along the line $k_x = 0$.  While a gapless $p_x$ superconducting phase is not our primary focus, we note that realizing such a state in a (110) quantum well with negligible Rashba coupling would be interesting in its own right.  

To gain a more complete picture of topological superconductor's stability in the physically relevant regime, we further illustrate the behavior of the bulk excitation gap in Figs.\ \ref{GapFig110}(b) and (c), fixing for concreteness $\gamma = 1$ and $m\lambda_D^2/V_y = 2$.  Figure \ref{GapFig110}(b) plots the dependence of the gap on $\Delta/V_y$ and $\lambda_R/\lambda_D$ when $\mu = 0$, while Fig.\ \ref{GapFig110}(c) displays the gap as a function of $\lambda_R/\lambda_D$ and $\mu/V_y$ when $\Delta/V_y = 0.66$.  

\subsection{Experimental considerations for (110) quantum wells}

The main drawback of our proposal compared to the Rashba-only model discussed by Sau \emph{et al}.\ can be seen in Fig.\ \ref{GapFig110}(b).  In the previous section, we discussed that it may be desirable to intentionally suppress the gap for the topological superconducting state by considering Zeeman splittings which greatly exceed the Rashba energy scale $m\alpha^2$, in order to achieve higher densities and thereby reduce disorder effects.  Here, however, this is possible to a lesser extent since the desired strength of $V_y$ is limited by the induced pairing field $\Delta$.  If $\Delta/V_y$ becomes too small, then the system enters the gapless regime as shown in Fig.\ \ref{GapFig110}(b). 

Nevertheless, our proposal has a number of virtues, such as its tunability.  As in the proposal of Sau \emph{et al}.\cite{Sau}, the strength of Rashba coupling can be controlled by applying a gate voltage\cite{RashbaTuning}, and the chemical potential in the semiconductor can be independently tuned by changing the Fermi level in the proximate $s$-wave superconductor.  In our case the parameter $\gamma\propto \sqrt{m_x/m_y}$ can be controlled to some extent by applying pressure to modify the mass ratio $m_x/m_y$, although this is not essential.  More importantly, one has additional control over the Zeeman field, which is generated by an externally applied in-plane magnetic field that largely avoids unwanted orbital effects.  Such control enables one to readily tune the system across the quantum phase transition separating the ordinary and topological superconducting phases [see Fig.\ \ref{GapFig110}(b)].  This feature not only opens up the opportunity to study this quantum phase transition experimentally, but also provides an unambiguous diagnostic for identifying the topological phase.  For example, the value of the critical current in the quantum well should exhibit a singularity at the phase transition, which would provide one signature for the onset of the topological superconducting state.  We also emphasize that realizing the required Zeeman splitting through an applied field is technologically far simpler than coupling the quantum well to a ferromagnetic insulator, and avoids the additional source of disorder generated by doping the quantum well with magnetic impurities.  

Since the extent to which one can enhance the density in the topological superconducting phase by applying large Zeeman fields is limited here, it is crucial to employ materials with appreciable Dresselhaus coupling.  We suggest that InSb quantum wells may be suitable for this purpose.  Bulk InSb enjoys quite large Dresselhaus spin-orbit interactions of strength 760eV\AA$^3$ (for comparison, the value in bulk GaAs is 28eV\AA$^3$; see Ref.\ \onlinecite{WinklerBook}).  For a quantum well of width $w$, one can crudely estimate the Dresselhaus coupling to be $\lambda_D\sim 760$eV\AA$^3/w^2$; assuming $w = 50$\AA, this yields a sizable $\lambda_D\sim 0.3$eV\AA.  Bulk InSb also exhibits a spin-orbit enhanced $g$-factor of roughly 50 (though confinement effects can substantially diminish this value in a quantum well\cite{WinklerBook}).  The large $g$-factor has important benefits.  For one, it ensures that Zeeman energies $V_y$ of order a Kelvin, which we presume is the relevant scale for $\Delta$, can be achieved with fields substantially smaller than a Tesla.  The ability to produce Zeeman energies of this scale with relatively small fields should open up a broad window where $V_y$ exceeds $\Delta$ but the applied field is smaller than the critical field for the proximate $s$-wave superconductor (which can easily exceed 1T).   Both conditions are required for realizing the topological superconducting state in our proposed setup.  A related benefit is that the Zeeman field felt by the semiconductor will be significantly larger than in the $s$-wave superconductor, since the $g$-factor for the latter should be much smaller.  This further suggests that $s$-wave superconductivity should therefore be disturbed relatively little by the required in-plane fields.

\section{Discussion}

Amongst the proposals noted in the introduction, the prospect for realizing Majorana fermions in a semiconductor sandwiched between a ferromagnetic insulator and $s$-wave superconductor stands out in part because it involves rather conventional ingredients (semiconductor technology is extraordinarily well developed).  Nevertheless, this setup is not without experimental challenges, as we attempted to highlight in Sec.\ II above.  For instance, a good interface between a ferromagnetic insulator and the semiconductor is essential, which poses an important engineering problem.  If one employs a magnetic semiconductor instead, this introduces an additional source of disorder (in any case magnetic semiconductors are typically hole doped).

The main goal of this paper was to simplify this setup even further, with the hope of hastening the experimental realization of Majorana fermions in semiconductor devices.  We showed that a topological superconducting state can be driven by applying a (relatively weak) in-plane magnetic field to a (110) semiconductor quantum well coupled \emph{only} to an $s$-wave superconductor.  The key to realizing the topological phase here was an interplay between Dresselhaus and Rashba couplings; together, they cause the spins to orient within a plane which tilts away from the quantum well.  An in-plane magnetic field then plays the same role as the ferromagnetic insulator or an applied perpendicular magnetic field plays in the Rashba-only models considered in Refs.\ \onlinecite{FujimotoTSC,SatoFujimoto,Sau}, but importantly without the detrimental orbital effects of the perpendicular field.  This setup has the virtue of simplicity---eliminating the need for a proximate ferromagnetic insulator or magnetic impurities---as well as tunability.  Having control over the Zeeman field allows one to, for instance, readily sweep across the quantum phase transition from the ordinary to the topological superconducting state.  Apart from fundamental interest, this phase transition can serve as a diagnostic for unambiguously identifying the topological phase experimentally (\emph{e.g.}, through critical current measurements).  As a more direct probe of Majorana fermions, a particularly simple proposal for their detection on the surface of a topological insulator was recently put forth by Law, Lee, and Ng\cite{MajoranaDetection}.  This idea relies on `Majorana induced resonant Andreev reflection' at a chiral edge.  In a topological insulator, such an edge exists between a proximity-induced superconducting region and a ferromagnet-induced gapped region of the surface.  In our setup, this effect can be realized even more simply, since the semiconductor will exhibit a chiral Majorana edge at its boundary, without the need for a ferromagnet.  Finally, since only one side of the semiconductor need be contacted to the $s$-wave superconductor, in principle this leaves open the opportunity to probe the quantum well directly from the other.  

The main disadvantage of our proposal is that if the Zeeman field in the semiconductor becomes too large compared to the proximity-induced pair field $\Delta$, the topological phase gets destroyed [see Fig.\ \ref{GapFig110}(b)].  By contrast, in the setup proposed by Sau \emph{et al}., the topological superconductor survives even when the Zeeman field greatly exceeds both $\Delta$ and $m\alpha^2$.  Indeed, we argued that this regime is where experimentalists may wish to aim, at least initially, if this setup is pursued.  Although the gap in the topological phase is somewhat suppressed in the limit $m\alpha^2/V_z \ll 1$, large Zeeman fields allow the density in this phase to be increased by one or two orders of magnitude, thus reducing disorder effects.  (We should note, however, that the actual size of Zeeman fields that can be generated by proximity to a ferromagnetic insulator or intrinsic magnetic impurities is uncertain at present.)  Since one is not afforded this luxury in our (110) quantum well setup, it is essential to employ materials with large Dresselhaus spin-orbit coupling in order to achieve reasonable densities in the semiconductor.  We argued that fairly narrow InSb quantum wells may be well-suited for this purpose.  Apart from exhibiting large spin-orbit coupling, InSb also enjoys a large $g$-factor, which should allow for weak fields (much less than 1T) to drive the topological phase in the quantum well while disturbing the proximate $s$-wave superconductor relatively little.  

There are a number of open questions which are worth exploring to further guide experimental effort in this direction.  As an example, it would be worthwhile to carry out more accurate modeling, including for instance cubic Rashba and Dresselhaus terms\cite{SpinOrbitHigherOrder} and (especially) disorder, to obtain a more quantitative phase diagram for either of the setups discussed here.  Exploring the full spectrum of vortex bound states (beyond just the zero-energy Majorana mode) is another important problem.  The associated `mini-gap'  provides one important factor determining the feasibility of quantum computation with such devices.  We also think it is useful to explore other means of generating topological superconducting phases in such semiconductor settings.  One intriguing possibility would be employing nuclear spins to produce a Zeeman field in the semiconductor\footnote{We thank Gil Refael and Jim Eisenstein for independently suggesting this possibility.}.  More broadly, the proposals considered here can be viewed as examples of a rather general idea discussed recently\cite{Shinsei} for eliminating the so-called fermion-doubling problem that can otherwise destroy the non-Abelian statistics\cite{Doron} necessary for topological quantum computation.  Very likely, we have by no means exhausted the possible settings in which Majorana fermions can emerge, even within the restricted case of semiconductor devices.  Might hole-doped semiconductors be exploited in similar ways to generate topological superconducting phases, for instance, or perhaps heavy-element thin films such as bismuth?

\acknowledgments{It is a pleasure to acknowledge a number of stimulating discussions with D.\ Bergman, S.\ Das Sarma, J.\ Eisenstein, M.\ P.\ A.\ Fisher, S.\ Fujimoto, R.\ Lutchyn, O.\ Motrunich, G.\ Refael, J.\ D.\ Sau, and A.\ Stern.  We also acknowledge support from the 
Lee A.\ DuBridge Foundation.

%\bibliography{Majorana110}

\end{document}